\documentclass[letterpaper]{aamas2009}
\usepackage{epsfig,graphicx}	
\newtheorem{theorem}{Theorem}[section]
\newtheorem{corollary}[theorem]{Corollary}
\newtheorem{lemma}[theorem]{Lemma}
\newtheorem{proposition}[theorem]{Proposition}
\newtheorem{definition}[theorem]{Definition}
\newtheorem{non-theorem}{Non-theorem}[section]

\newtheorem{assumption}{Assumption}[section]
\newcommand{\assum}{\begin{assumption}}
\newcommand{\eassum}{\end{assumption}}
\newcommand{\thm}{\begin{theorem}}
\newcommand{\lem}{\begin{lemma}}
\newcommand{\pro}{\begin{proposition}}
\newcommand{\dfn}{\begin{definition} \rm}
\newcommand{\rem}{\begin{remark}}
\newcommand{\xam}{\begin{example}}
\newcommand{\cor}{\begin{corollary}}
\newcommand{\prf}{\begin{proof}}
\newcommand{\ethm}{\end{theorem}}
\newcommand{\elem}{\end{lemma}}
\newcommand{\epro}{\end{proposition}}
\newcommand{\edfn}{\bbox\end{definition}}
\newcommand{\erem}{\bbox\end{remark}}
\newcommand{\exam}{\bbox\end{example}}
\newcommand{\ecor}{\end{corollary}}
\newcommand{\eprf}{\end{proof}}
\newcommand{\beqn}{\begin{equation}}
\newcommand{\eeqn}{\end{equation}}
\newcommand{\bbox}{\vrule height7pt width4pt depth1pt}
\newcommand{\commentout}[1]{}

\newenvironment{RETHM}[2]{\trivlist \item[\hskip 10pt\hskip\labelsep{\sc #1\hskip 5pt\relax\ref{#2}.}]\it}{\endtrivlist}
\newcommand{\rethm}[1]{\begin{RETHM}{Theorem}{#1}}
\newcommand{\repro}[1]{\begin{RETHM}{Proposition}{#1}}
\newcommand{\relem}[1]{\begin{RETHM}{Lemma}{#1}}
\newcommand{\recor}[1]{\begin{RETHM}{Corollary}{#1}}

\newcommand{\erethm}{\end{RETHM}}
\newcommand{\erepro}{\end{RETHM}}
\newcommand{\erelem}{\end{RETHM}}
\newcommand{\erecor}{\end{RETHM}}

\newcommand{\ABR}{\mathit{ABR}}

\DeclareMathOperator*{\argmax}{argmax}

\begin{document}

\title{Multiagent Learning in Large Anonymous Games}

\numberofauthors{3}
\author{
\alignauthor Ian A. Kash\\
\affaddr{Computer Science Dept.}\\
\affaddr{Cornell University}\\
\email{kash@cs.cornell.edu}
\alignauthor Eric J. Friedman\\
\affaddr{School of Operations Research and Information Engineering}\\
\affaddr{Cornell University}\\
\email{ejf27@cornell.edu}
\alignauthor Joseph Y. Halpern\\
\affaddr{Computer Science Dept.}\\
\affaddr{Cornell University}\\
\email{halpern@cs.cornell.edu}
}

\maketitle

\begin{abstract}

In large systems, it is important for agents to learn to act
effectively, but sophisticated multi-agent learning algorithms
generally do not scale.
An alternative approach is to find restricted classes of games where
simple, efficient algorithms converge.
It is shown that stage learning efficiently converges to Nash
equilibria in large anonymous games if best-reply dynamics converge.
Two features are identified that
improve convergence. First, rather than making learning more
difficult, more agents are actually beneficial in many settings.
Second, providing agents with statistical information about the
behavior of others can significantly reduce the number of
observations needed.
\end{abstract}




\section{Introduction}\label{sec:intro}

Designers of distributed systems are frequently unable to
determine how an agent in the system should behave, because
optimal behavior depends on the user's preferences and the
actions of others.  A natural approach is to have agents use a
learning algorithm.
Many multiagent learning algorithms have been proposed including
simple strategy update procedures such as
\emph{fictitious play}~\cite{fudenbergandlevine}, multiagent versions of
\emph{Q-learning}~\cite{watkins92}, and \emph{no-regret
algorithms}~\cite{lugositext}.

However, as we discuss in Section~\ref{sec:relatedwork}, existing
algorithms are generally unsuitable for large distributed systems.  In
a distributed system, each agent has a limited view of the
actions of other agents.  Algorithms that require knowing, for
example, the strategy chosen by every agent cannot
be implemented.  Furthermore, the size of distributed systems
requires fast convergence.  Users may use the system for short
periods of time and conditions in the system change over time, so
a practical algorithm for a system with thousands or millions of users
needs to have a convergence rate that is sublinear in the number of
agents.
Existing algorithms tend
to provide performance guarantees that are polynomial or even
exponential.  Finally, the large number of agents in the system
guarantees that there will be noise.  Agents will make mistakes and
will behave in unexpectedly.
Even if no agent changes his strategy,
there can still be noise in agent payoffs.  For example, a gossip
protocol will match different agents from round to round; congestion
in the underlying network may effect message delays between agents.
A learning algorithm needs to be robust to this noise.

While finding an algorithm that satisfies these requirements for
arbitrary games may be difficult,
distributed systems have
characteristics that make the problem easier.
First, they involve a large number of agents.
Having more agents may seem to make learning harder---after all,
there are more possible interactions.  However, it has the advantage that
the 
outcome of an action typically depends
only weakly on what other agents do.  This makes outcomes
robust to noise.
Having a large number of agents also make it less useful for an agent
to  try to influence others; it becomes a better policy
to try to learn an optimal response.  In contrast, with a
small number of agents, an agent can attempt to
guide learning agents into an outcome that is beneficial for him.

Second,
distributed systems
are often \emph{anonymous}~\cite{blonski01};
it does not matter {\em who} does something, but rather
{\em how many} agents do it.  For example, when there is congestion on a
link, the experience of a single agent does not depend on who is
sending the packets, but on how many are being sent.

Finally, and perhaps most importantly, in a distributed system the
system designer controls the game agents are playing.  This gives us a
somewhat different perspective than most work, which takes the
game as
given.  We do not need to solve the hard problem of finding an
efficient algorithm for all games.  Instead, we can find
algorithms that work efficiently for interesting classes of games,
where for us ``interesting'' means ``the type of games a system designer
might wish agents to play.''  
Such games should 
be ``well behaved,'' since it would be strange to design a 
system where an agent's decisions can influence other agents in pathological
ways.

In Section~\ref{sec:theory}, we show that
{\em stage learning}~\cite{friedman98} is robust,
implementable with minimal information, and converges efficiently for
an interesting class of games.  In this algorithm, agents
divide the rounds of the game into a series of stages.  In each stage,
the agent uses a fixed strategy except that he occasionally explores.
At the end of a stage, the agent chooses as his strategy for the next
stage whatever strategy had the highest average reward in the current
stage.
We prove that, under appropriate conditions, a large system of stage
learners will follow (approximate) best-reply dynamics despite errors and
exploration.

For games where best-reply dynamics converge, our theorem
guarantees
that learners will play an approximate Nash equilibrium.
In contrast to previous results where the convergence guarantee scales
poorly with the number of agents, our theorem guarantees convergence
in a finite amount of time with an infinite number of agents.
While the assumption that best-reply dynamics converge is
a strong one,
many interesting games converge under best-reply dynamics,
including dominance solvable games and games with monotone best
replies.
Marden et al.~\citeyear{marden07a}
have observed that convergence of best-reply dynamics
is often a property of games that humans design.
Moreover, convergence of best-reply dynamics is a weaker assumption
than
a common assumption
made in the mechanism design
literature, that the games of interest
have dominant strategies (each agent has a strategy that is optimal no
matter what other agents do).

Simulation results, presented
in Section~\ref{sec:empirical}, show that convergence is fast in
practice: a system with thousands of agents can converge in a few
thousand rounds.  Furthermore, we identify two factors that
determine the rate and quality of convergence.  One is the number of
agents: having more agents makes the noise in the systen more
consistent so agents can learn using fewer observations.
The other is giving agents statistical information about the
behavior of other agents; this can speed convergence by an order of
magnitude.
Indeed, even noisy statistical information about agent behavior, which
should be relatively easy to obtain and disseminate,  can
significantly improve performance.

\section{Related Work}
\label{sec:relatedwork}

One approach to learning to play games is to generalize
reinforcement learning algorithms such as
Q-learning~\cite{watkins92}.  One nice feature of this
approach is that it can handle games with state, which is
important in distributed systems.  In Q-learning, an agent associates
a value with each state-action pair.  When he chooses action $a$ in
state $s$, he updates the value $Q(s,a)$ based on the reward he
received and the best value he can achieve in the resulting state $s'$
($\max_{a'} Q(s',a')$).  When generalizing to multiple agents, $s$ and
$a$ become vectors of the state and action of every agent and
the $\max$ is replaced by a prediction of the behavior of other
agents.  Different algorithms use different predictions; for example,
Nash-Q uses a Nash equilibrium calculation~\cite{nashq}.
See~\cite{shoham03} for a survey.

Unfortunately, these algorithms converge too slowly
for a large distributed system.
The algorithm needs to experience each possible action profile
many times to guarantee convergence.
So, with $n$ agents and $k$ strategies, the naive convergence time is
$O(k^n$).  Even with a
better representation for anonymous games, the convergence time is
still $O(n^k)$ (typically $k \ll n$).
There is also a  more fundamental problem with this approach: it
assumes information that an agent is unlikely to have. In order to know
which value to update, the agent must learn the action chosen by every
other agent. In practice, an agent will learn
something about the actions of the agents with whom he directly
interacts, but is unlikely to gain much information about
the actions of other agents.

Another approach is
\emph{no-regret learning}, where agents choose a strategy for each
  round
that guarantees 
that
the
regret of their choices will be low.
  Hart and Mas-Colell~\cite{hart00}
present such a learning
procedure that converges to a \emph{correlated equilibrium} \cite{OR94}
given knowledge of what the payoffs of every action
would have been in each round. They also provide a variant of their
algorithm that requires only information about the 
agent's
actual payoffs~\cite{hart01}. However, to guarantee convergence to
within $\epsilon$ of a correlated equilibrium requires $O(kn /
\epsilon^2 \log kn)$, still too slow for large systems.
Furthermore, the convergence guarantee is that the distribution of
play converges to equilibrium; the strategies of individual learners
will not converge.
Better results can be achieved in restricted settings.  For example,
Blum et al.~\cite{blum06} showed that in routing games a
continuum of no-regret learners will approximate Nash equilibrium in a
finite amount of time.

Foster and Young~\cite{young06} use a stage-learning procedure that
converges to Nash equilibrium for two-player
games.  Germano
and Lugosi~\cite{germano07} showed that it converges for
generic $n$-player games (games where best replies are unique).
Young~\cite{young08} uses a similar algorithm without explicit stages
that also converges for generic $n$-player games.
Rather than selecting best replies, in these algorithms
agents choose new actions randomly when not in equilibrium.
Unfortunately, these algorithms involve searching the whole strategy
space, so their convergence time is exponential.
Another algorithm that uses stages to provide
a stable learning environment is the ESRL algorithm for
coordinated exploration~\cite{verbeeck07}.

Marden et al.~\cite{marden07b,marden08} use an algorithm with
experimentation and best replies but without explicit stages that
converges for \emph{weakly acyclic games}, 
where best-reply dynamics converge 
when agents move one at a time, rather than moving all at once, as we
  assume here.
Convergence is based on the existence of a sequence of exploration
  moves that lead to equilibrium.  With $n$ agents who explore with
  probability $\epsilon$, this analysis gives a convergence time of
  $O(1/\epsilon^n)$.  Furthermore, the guarantee requires $\epsilon$
to be sufficiently small that agents essentially explore one at a
time, so
  $\epsilon$ needs to be $O(1/n)$.

There is a long history of
work examining simple learning procedures such as
\emph{fictitious play}~\cite{fudenbergandlevine}, where each
agent makes a best response assuming that each other player's strategy is
characterized by the empirical frequency of his observed moves.
In contrast to algorithms with convergence guarantees for general
games, these algorithms fail to converge in many games.
But for classes of games where
they do converge, they tend to do so rapidly.
However, most
work in this area assumes that the actions of agents are
observed by all agents, agents know the
payoff matrix, and payoffs are deterministic.
A recent approach in this tradition is based on the Win or Learn Fast
principle, which has limited convergence guarantees but often performs
well in practice~\cite{wolf}.

There is also a body of empirical work on the convergence of
learning algorithms in multiagent settings.  Q-learning
has had empirical success in 
pricing games~\cite{tesauro02}, $n$-player cooperative games~\cite{claus98},
and grid world games~\cite{bowling00}.
Greenwald at al.~\cite{greenwald01} showed that a number of
algorithms, including stage learning, converge in a variety of simple
games.  Marden et al.~\cite{marden08} found that their algorithm
converged 
must faster in a congestion game than
the theoretical analysis would suggest.
Our theorem suggests 
an explanation for these empirical
observations:
best-reply dynamics converge in all these games.
While our theorem applies directly only to stage learning, it provides
intuition as to why algorithms that learn ``quickly enough''
and change their behavior ``slowly enough'' rapidly converge to Nash
equilibrium in practice.

\section{Theoretical Results} \label{sec:theory}

\subsection{Large Anonymous Games}

We are interested in anonymous games with countably many agents.
Assuming that there are countably many agents simplifies the proofs; it
is straightforward to extend our results to
games with a large finite number of agents.
Our model is adapted from that of~\cite{blonski01}.
Formally, a \emph{large anonymous game} is
characterized by a tuple $\Gamma = (\mathbb{N},A,P,\Pr)$.
\begin{itemize}

\item $\mathbb{N}$ is the countably infinite set of agents.

\item $A$ is a finite set of actions from which each agent can
choose (for simplicity, we assume that each agent can choose from the
same set of actions).

\item $\Delta(A)$, the set of probability distributions over $A$,
has two useful interpretations.  The first is as
the set of mixed actions.  For $a \in A$ we will
abuse notation and denote the mixed
action that is $a$ with probability 1 as $a$.  In each round each
agent chooses one of these mixed actions.
The second interpretation of $\rho \in \Delta(A)$ is as the fraction of
agents choosing each action $a \in A$.
This is important for our notion of anonymity, which says an agent's
utility should depend only on how many agents choose each action
rather than who chooses it.

\item $G = \{ g : \mathbb{N} \rightarrow \Delta(A) \}$
is the set of (mixed) action profiles
(i.e. which action each agent chooses).
Given the mixed action of every agent, we want to know the
fraction of agents that end up choosing action $a$.
For $g \in G$, let $g(i)(a)$ denote the probability with which agent $i$
plays $a$ according to $g(i) \in \Delta(A)$.  We can then express the
fraction of agents in $g$ that choose action $a$ as
$\lim_{n \rightarrow \infty} (1 / n) \sum_{i = 0}^n g(i)(a)$, if this
limit exists.
If the limit exists for all actions $a \in A$, let $\rho_g \in
\Delta(A)$ give the value of the limit for each $a$.
The profiles $g$ that we use are all determined by a simple random
process.  For such profiles $g$,
the strong law of large numbers
(SLLN)
guarantees that with probability 1
$\rho_g$ is well defined.
Thus
it will typically be well defined (using similar limits) for us to
talk about the fraction of agents who do something.

\item $P \subset \mathbb{R}$ is a finite set of payoffs agents can
receive.

\item $\Pr: A \times \Delta(A) \rightarrow \Delta(P)$ denotes the
distribution over payoffs that results when the agent
performs action $a$ and other agents follow action profile $\rho$.
We use a probability distribution over payoffs rather than a payoff
to model the fact that agent payoffs may change even if no
agent changes his strategy.
The expected utility of an agent
who performs mixed action $s$ when other agents follow action
distribution $\rho$ is
$u(s,\rho) = \sum_{a \in A} \sum_{p \in P} p s(a)
\Pr_{a,\rho}(p)$.
Our definition of $\Pr$ in terms of $\Delta(A)$ rather than $G$
ensures the the game is anonymous.  We further
require that $\Pr$ (and thus $u$) be \emph{Lipschitz continuous}.%
\footnote{Lipschitz continuity imposes the additional constraint that
  there is some constant $K$ such that $|\Pr(a,\rho) - \Pr(a,\rho')| /
  ||\rho - \rho'||_1 \leq K$ for all $\rho$ and $\rho'$.  Intuitively,
  this ensures that the distribution of outcomes doesn't change ``too
  fast.''
This is a standard assumption that is easily seen to hold
in the games that have typically been considered in the literature.}
  For definiteness, we use the L1 norm as our notion
of distance when specifying continuity (the L1 distance between two
vectors is the sum of the absolute values of the differences in each
component).
Note that
this formulation assumes all agents share a common utility
function.

\end{itemize}

An example of a large anonymous game  is one where, in each
round, each agent plays a two-player game against an opponent
chosen at random.
Then $A$ is the set of actions of the two-player
game and $P$ is the set of payoffs of the game.  Once every agent
chooses an action, the distribution over actions is characterized by
some $\rho \in \Delta(A)$.
Let $p_{a,a'}$ denote the payoff for the
agent if he plays $a$ and the other agent plays $a'$.
Then the utility of mixed action $s$ given distribution $\rho$ is
$$u(s,\rho) = \sum_{a,a' \in A^2} s(a) \rho(a') p_{a,a'}.$$

\subsection{Best-Reply Dynamics} \label{ssec:br}

Given a game $\Gamma$ and an action distribution $\rho$, a natural
goal for an agent is to play the action that maximizes his expected
utility with respect to $\rho$: $\argmax_{a \in A} u(a,\rho)$.  We
call such an action a {\em best reply} to $\rho$.
In a practical amount of time, an agent may have difficulty
determining which of two actions with close expected utilities is
better, so we will allow agents to choose actions that are close to
best replies.  If $a$ is a best reply to $\rho$, then $a'$ is an
{\em $\eta$-best reply} to $\rho$ if $u(a',\rho) + \eta
  \geq u(a,\rho)$.  There may be more than one $\eta$-best reply; we
  denote the set of $\eta$-best replies $\ABR_\eta(\rho)$.

We do not have a single agent looking for a best reply;
every agent is trying to find a one at the same time.
If agents start off with some action distribution $\rho_0$, after
they all find a best reply there will be a new action distribution
$\rho_1$.
We assume
that
$\rho_0(a) = 1 / |A|$ (agents choose their initial strategy
uniformly at random), but our results apply to any
distribution used to determine the initial strategy.
We say that a sequence $(\rho_0, \rho_1, \ldots)$ is an
\emph{$\eta$-best-reply sequence} if the support of $\rho_{i+1}$
is a subset of $\ABR_\eta(\rho_i)$; that is $\rho_{i+1}$ gives
positive probability only to approximate best replies to $\rho_i$.
A $\eta$ best-reply sequence {\em converges} if there exists
some $t$ such that for all $t' > t$, $\rho_{t'} = \rho_t$.
Note that this is a particularly strong notion of convergence because
we require the $\rho_t$ to converge in finite time and not merely in
the limit.
A game may have infinitely many best-reply sequences, so we say that
{\em approximate best-reply dynamics converge} if there exists
some $\eta > 0$ such that every $\eta$-best-reply sequence
converges.
The limit distribution $\rho_t$ determines a mixed strategy that is an
$\eta$-Nash equilibrium.

Our theorem shows that learners can
successfully learn in large anonymous games
where approximate best-reply dynamics converge.
The number of stages needed to converge is determined by
the number of best replies needed before the sequence converges.  It
is possibly to design games that have long best-reply sequences, but
it practice most games have short sequences.
One condition that guarantees this is if $\rho_0$ and all
the degenerate action distributions $a \in A$
(i.e., distributions that assign probability 1 to some $a \in A$)
have unique best replies. In this case, there can be
at most $|A|$ best replies before equilibrium is reached.
Furthermore, in such games the distinction between $\eta$-best
replies and best replies is irrelevant; for sufficiently small
$\eta$, a $\eta$-best reply is a best reply.
It is not hard to show that the property that degenerate strategies
have unique best replies is generic; it holds for almost every game.

\subsection{Stage Learners} \label{sec:stage}

An agent who wants to find a best reply
may not know the set of payoffs $P$, the mapping from actions to
distributions over payoffs $\Pr$, or the action distribution
$\rho$ (and, indeed, $\rho$  may be changing over time), so he
will have to use some type of learning algorithm to learn it.
Our approach is to divide the
play of the game into a sequence of stages.  In each stage, the agent
almost always plays some fixed action $a$, but also explores other
actions.  At the end of the stage, he chooses a new $a'$ for the next
stage based on what he has learned.  An important feature of this
approach is that agents maintain their actions for the entire
stage, so each stage provides a stable environment in which agents can
learn.  To simplify our results, we specify a way of
exploring and learning within a stage (originally described
in~\cite{friedman98}), but our results
should generalize to any ``reasonable'' learning algorithm used to
learn within a stage.
(We discuss what is ``reasonable'' in Section~\ref{sec:discussion}.)
In this section, we show
that, given a suitable parameter, at the each stage most agents will
have learned a best reply to the environment of that stage.

Given a game $\Gamma$, in each round $t$ agent $i$ needs to select a
mixed action $s_{i,t}$.  Our agents use strategies
that we denote $a_\epsilon$, for $a \in A$, where
$a_\epsilon(a) = 1 - \epsilon$ and
$a_\epsilon(a' \neq a) = \epsilon / (|A| - 1)$.
Thus, with $a_\epsilon$, an agent almost always plays $a$, but with
probability $\epsilon$
explores
other strategies uniformly at random.
Thus far we have not specified what information an agent can use to
choose $s_{i,t}$.
Different games may provide different
information.  All that we require is that an agent know all of his
previous actions and his previous payoffs.  More precisely, for all
$t' < t$, he knows his action $a_{t'}(i)$ (which is determined by
$s_{i,t'}$) and his payoffs $p_{t'}(i)$ (which is determined
by $\Pr(a_{i,t'},\rho_{t'})$, where $\rho_{t'}$ is the
action distribution for round $t'$; note that we do not assume that
the agent knows $\rho_{t'}$.)
Using this information, we can
express the average value of an action over the previous
$\tau = \lceil 1 / \epsilon^2 \rceil$ rounds (the length of a stage).%
\footnote{The use of the exponent 2 is arbitrary.  We require only
that the expected number of times a strategy is explored increases
as $\epsilon$ decreases.}
Let
$H(a,i,t) = \{ t - \tau \leq t' < t ~|~ a_{t'}(i) = a \}$ be the set
of recent rounds in which $a$ was played by $i$.  Then the average
value is $V(a,i,t) = \sum_{t' \in H(a,i,t)} p_{t'}(i) / |H(a,i,t)|$
if $|H(a,i,t)| > 0$ and 0 otherwise. While we need the value of $H$
only at times that are multiples of $\tau$, for convenience we
define it for arbitrary times $t$.

We say that an agent is an {\em $\epsilon$-stage learner} if he
chooses his actions as follows.  If $t = 0$, $s_t$ is chosen
at random from $\{ a_\epsilon ~|~ a \in A \}$.
If $t$ is a nonzero multiple of $\tau$,
$s_{i,t} = a(i,t)_\epsilon$ where
$a(i,t) = \argmax_{a \in A} V(a,i,t)$.
Otherwise, $s_{i,t} = s_{i,t-1}$.  Thus, within a stage, his mixed
action is fixed and at the end of a stage he updates it to
use the action with the highest average value during the previous
stage.

The evolution of a game played by stage learners is not deterministic;
each agent chooses a random $s_{i,0}$ and the sequence of $a_t(i)$
and $p_t(i)$ he observes is also random.   However, with a countably
infinite set of agents, we can use the SLLN to
make statements about the overall behavior of the game.
Let $g_t(i) = s_{i,t}$.  A {\em run} of the game consists of a
sequence of triples $(g_t,a_t,p_t)$.  The
SLLN guarantees that with probability 1 the
fraction of agents who
choose a strategy $a$ in $a_t$ is $\rho_{g_t}(a)$.  Similarly, the
fraction of agents who chose $a$ in $a_t$ that receive payoff $p$ will
be $\Pr(a,\rho_{g_t})(p)$ with probability 1.

To make our notion of a stage precise, we refer to
the sequence of tuples $(g_{n\tau},a_{n\tau},p_{n\tau}) \ldots
(g_{(n+1)\tau - 1},a_{(n+1)\tau - 1},p_{(n+1)\tau - 1})$ as stage $n$
of the run.  During stage $n$ there is a stationary action
distribution that we denote $\rho_{g_{n\tau}}$.
If $s_{i,(n+1)\tau} = a_\epsilon$ and
$a \in \ABR_\eta(g_{n\tau})$, then we say that agent $i$ has
{\em learned an $\eta$-best reply}
during stage $n$ of the run.  As the following lemma
shows, for sufficiently small $\epsilon$, most agents will learn an
$\eta$-best reply.

\lem \label{lem:stage}
For all large anonymous games $\Gamma$, action profiles,
approximations $\eta > 0$, and probabilities of error $e > 0$,
there exists an $\epsilon^* > 0$ such that
for $\epsilon < \epsilon^*$ and all $n$,
if all agents are $\epsilon$-stage learners, then
at least a $1 - e$ fraction of agents will learn an $\eta$-best reply
during stage $n$.
\elem

\prf
(Sketch)
On average, an agent using strategy $a_\epsilon$ plays action $a$
$(1-\epsilon)\tau$ times
during a stage and plays all other actions $\epsilon
\tau/(n-1)$ times each. For $\tau$ large, the realized
number of times played will be close to the expectation value with
high probability.  Thus, if $\epsilon \tau$ is sufficiently large,
then the average payoff from each action will be exponentially
close to the true expected value (via a standard Hoeffding bound
on sums of i.i.d.~random variables), and thus each
the learner will correctly identify an action with approximately the
highest expected payoff with probability at least $1 - e$.
By the
SLLN, at least a $1-e$ fraction of agents will
learn an $\eta$-best reply.
A detailed version of this proof in a more
general setting can be found in \cite{friedman98}.
\eprf

\subsection{Convergence Theorem}

Thus far we have defined large anonymous games
where approximate best-reply dynamics converge.
If all agents in the game are
$\epsilon$-stage learners, then the sequence
$\hat{\rho}_0, \hat{\rho}_1, \ldots$
of action distributions in a run of the game is not a
best-reply sequence, but it is close.  The
action used by most agents
most of the time in each $\hat{\rho}_n$ is the action used in $\rho_n$
for some approximate best reply sequence.

In order to prove this, we
need to define ``close.''  Our definition is
based on the error rate $e$ and exploration rate $\epsilon$ that
introduces noise into $\hat{\rho}_n$.  Intuitively, distribution
$\hat{\rho}$ is close to $\rho$ if, by changing the strategies of
an $e$ fraction of agents and having all agents explore an $\epsilon$
fraction of the time, we can go from an action profile with
corresponding action distribution $\rho$ to one with corresponding
distribution $\hat{\rho}$.  Note that this definition will not be
symmetric.

In this definition, $g$ identifies what (pure) action each agent is
using that leads to $\rho$, $g'$ allows an $e$ fraction of agents to
use some other action,
and $\hat{g}$ incorporates the fact
that each agent is exploring, so each strategy is an $a_\epsilon$ (the
agent usually plays $a$ but explores with probability $\epsilon$).

\dfn \label{def:close}
Action distribution $\hat{\rho}$ {\em
$(e,\epsilon)$-close} to $\rho$ if there exist
$g$, $g'$, and $\hat{g} \in G$ such that:
\begin{itemize}

\item
$\rho = \rho_g$ and $\hat{\rho} = \rho_{\hat{g}}$;

\item
$g(i) \in A$ for all $i \in \mathbb{N}$;

\item
$||\rho_g - \rho_{g'}||_1 \leq 2e$
(this allows an $e$ fraction of agents in $g'$ to play a different strategy
from $g$);

\item
for some $\epsilon' \le \epsilon$,
if $g'(i) = a$ then $\hat{g}(i) = a_{\epsilon'}$.
\bbox
\end{itemize}
\end{definition}

The use of $\epsilon'$ in the final requirement ensures that if two
distributions are $(e,\epsilon)$-close then they are also
$(e',\epsilon')$-close for all $e' \geq e$
and $\epsilon' \geq \epsilon$.  As an example of the asymmetry of this
definition, $a_\epsilon$ is $(0,\epsilon)$ close to $a$, but the
reverse is not true.
While $(e,\epsilon)$-closeness is a useful distance measure
for our analysis, it is an unnatural notion of distance for specifying
the continuity of $u$, where we used the L1 norm.
The following simple lemma shows that this distinction is unimportant;
if $\hat{\rho}$ is sufficiently $(e,\epsilon)$-close to $\rho$ then
it is close according to the L1 measure as well.
\lem \label{lem:value}
If $\hat{\rho}$ is $(e,\epsilon)$-close to
$\rho$, then $||\hat{\rho} - \rho||_1 \leq 2(e + \epsilon)$.
\elem

\prf
Since $\hat{\rho}$ is
$(e,\epsilon)$-close to $\rho$, there exist $g$, $g'$, and $\hat{g}$
as in Definition~\ref{def:close}.
Consider the distributions $\rho_g
= \rho$, $\rho_{g'}$, and $\rho_{\hat{g}} = \hat{\rho}$.
We can view these three distributions as vectors, and calculate their L1
distances.
By Definition~\ref{def:close},
$||\rho_g - \rho_{g'}||_1 \leq 2e$.
$||\rho_{g'} - \rho_{\hat{g}}||_1 \leq 2\epsilon$
because an $\epsilon$ fraction of agents explore.
Thus by the triangle inequality, the L1 distance
between $\rho$ and $\hat{\rho}$ is at most $2(e + \epsilon)$.
\eprf

We have
assumed that approximate best reply sequences of $\rho_n$ converge,
but during a run of the game agents will actually be learning
approximate best replies to $\hat{\rho}_n$.  The following lemma shows
that this distinction does not matter if $\rho$ and $\hat{\rho}$ are
  sufficiently close.

\lem \label{cor:br}
For all $\eta$ there exists a $d_\eta$ such that if
$\hat{\rho}$ is $(e,\epsilon)$-close to $\rho$, $e > 0$, $\epsilon >
0$, and $e + \epsilon < d_\eta$ then
$\ABR_{(\eta / 2)}(\hat{\rho}) \subseteq
\ABR_\eta(\rho)$.
\elem

\prf
Let $K$ be the maximum of the Lipschitz constants for all
$u(a,\cdot)$ and $d_\eta = \eta / (8K)$.  Then for all
$\hat{\rho}$ that are $(e,\epsilon)$-close to $\rho$ and all $a$,
$|u(a,\hat{\rho} - u(a,\rho)| \leq ||\hat{\rho} - \rho||_1 K
\leq 2 \eta / (8K) K = \eta / 4$ by
Lemma~\ref{lem:value}.

Let $a \notin \ABR_\eta(\rho)$ and
$a' \in \argmax_{a' \in \ABR_\eta(\rho)} u(a',\hat{\rho})$.
Then $u(a,\rho) + \eta < u(a',\rho)$.  Combining this with the
above gives $u(a,\hat{\rho}) + \eta / 2 < u(a',\hat{\rho})$.
Thus $a \notin \ABR_{\eta / 2}(\hat{\rho})$.
\eprf

Lemmas~\ref{lem:stage} and \ref{cor:br} give requirements on
$(e,\epsilon)$.  In the statement of the theorem, we call
$(e,\epsilon)$ {\em $\eta$-acceptable} if they
satisfy the requirements of both lemmas for $\eta / 2$
and all $\eta$-best-reply sequences converge in $\Gamma$.
\thm \label{thm:nash}
Let $\Gamma$ be a large anonymous game
where approximate best-reply dynamics converge and let $(e,\epsilon)$
be $\eta$-acceptable for $\Gamma$.  If all agents are $\epsilon$-stage
learners then, for all runs, there exists an $\eta$-best-reply
sequence $\rho_0,\rho_1, \ldots$
such that in stage $n$ at least a $1 - e$ fraction will learn a
best reply to $\rho_n$ with probability 1.
\ethm

\prf
$\rho_0 = \hat{\rho_0}$, so $\hat{\rho_0}$ is $(e,\epsilon)$-close to
$\rho$.   Assume $\hat{\rho}_n$ is $(e,\epsilon)$-close to $\rho$.  By
Lemma~\ref{lem:stage} at least a $1 - e$ fraction
will learn a $\eta / 2$-best reply to $\hat{\rho}_n$.  By
Lemma~\ref{cor:br}, this is a $\eta$-best reply to $\rho_n$.
Thus $\hat{\rho}_{n+1}$ will be $(e,\epsilon)$-close to $\rho_{n+1}$.
\eprf

Theorem~\ref{thm:nash} guarantees that after a finite number of stages,
agents will be 
close to an approximate Nash equilibrium profile. Specifically, 
$\hat{\rho}_n$ will be
$(e,\epsilon)$-close 
to an $\eta$-Nash equilibrium profile $\rho_n$. 
Note that this means that $\hat{\rho}_n$ 
is actually
an $\eta'$-Nash equilibrium for a larger $\eta'$ that depends on
$\eta$,$e$,$\epsilon$, and the Lipschitz constant $K$.

Our three requirements for a practical learning algorithm were that it
require minimal information, converge quickly in a large system, and
be robust to noise.  Stage learning  requires only that an agent know his
own payoffs, so the first condition is satisfied.
Theorem~\ref{thm:nash} shows that it satisfies the other two
requirements.  Convergence is guaranteed in a finite number of stages.
While the number of stages depends
on the game, in Section~\ref{ssec:br} we argued that in many cases it
will be quite small.
Finally, robustness comes from tolerating an $e$
fraction of errors.    While in our proofs we assumed these errors
were due to learning, the analysis is the same if some of this noise
is from other sources such as churn (agents entering and leaving the
system)
or agents making errors.
We discuss this issue more in Section~\ref{sec:discussion}.

\section{Simulation Results}
\label{sec:empirical}

Theorem~\ref{thm:nash} guarantees convergence for a sufficiently
small exploration probability $\epsilon$, but decreasing $\epsilon$
also increases $\tau$, the length of a stage.  Increasing the length
of a stage means that agents take longer to reach equilibrium, so
for stage learning to be practical, $\epsilon$ needs to be
relatively large. To show that $\epsilon$ can be large in practice,
we tested populations of stage learners in a number of games where
best reply dynamics converge and experienced convergence with
$\epsilon$ between $0.01$ and $0.05$. This allows convergence within
a few thousand rounds in many games. While our theorem applies only to stage
learning, the analysis provides intuition as to why a reasonable
algorithm that changes slowly enough that other learners have a
chance to learn best replies should converge as well.  To test a
very different type of algorithm,
we also implemented the no-regret
learning algorithm of Hart and Mas-Collell~\cite{hart01}.  This
algorithm also quickly converged close to Nash equilibrium, although
in many games it did not converge as closely as stage learning.

Our theoretical results make two significant predictions about
factors that influence the rate of convergence.  Lemma~\ref{lem:stage}
tells us that the length of a stage is determined by the number
of times each strategy needs to be explored to get an accurate
estimate of its value.  Thus the amount of
information provided by each observation has a large effect on the
rate of convergence.  For example, in a random matching game, an
agents payoff provides information about the strategy of one other
agent.  On the other hand, if he receives his expected payoff for
being matched, a single observation provides information about the
entire distribution of strategies.  In the latter case the agent can
learn with many fewer observations.

A related prediction is that having more agents
will  lead to faster convergence, particularly in games where
payoffs are determined by the average behavior of other agents,
because variance in payoffs due to exploration and mistakes decreases
as the number of agents increases.  Our experimental results
illustrate both of these phenomena.

We tested the learning behavior of stage learners and no-regret
learners in a number of games, including prisoner's dilemma, a
climbing game~\cite{claus98},
the congestion game described in \cite{greenwald01} with
both ACP and serial mechanisms, and two
different contribution games (called a Diamond-type search model
in~\cite{MiR90}).  
We
implemented payoffs both by randomly matching players and by giving
each player what his expected payoff would have been had he been
randomly matched
(some payoffs were adjusted to make the games symmetric).
Results were similar across the different games, so we report only the
results for a contribution game.

In the contribution game, agents choose strategies from 0 to 19,
indicating how much effort they contribute to a collective
enterprise.  The value to an agent depends on how much he contributes,
as well as how much other agents contribute.
If he contributes $x$ and the contribution of the other agents is $y$,
then his utility is $2xy - c(x)$, where $c(0) = 0$, $c(1) = 1$, $c(x)
= (x-1)^2$ for $x \in 2,\ldots,8$ and $c(x) = x^2 + 2n$ for $x > 8$.
We considered two
versions of this game.  In the first, $y$ is determined by the average
strategy of the other agents.  In the second, $y$ is determined by
randomly matching the agent with another agent.

Our implementation of stage learners is as described in
Section~\ref{sec:stage}, with $\epsilon = 0.05$ when $y$ is determined
by the average and $\epsilon = 0.01$ when $y$ is determined by random
matching.
Rather than
taking the length of stage $\tau$ as $1 / \epsilon^2$, we set $\tau =
250$ and $2000$, respectively; this gives better performance.
Our implementation of no-regret learners is based on
that of Hart and Mas-Colell~\cite{hart01}, with improvements
suggested by Greenwald et al.~\cite{greenwald01}.

\begin{figure}
\centering
\includegraphics[height=2.25in]{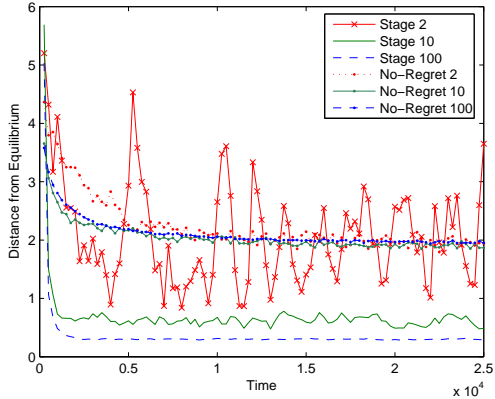}
\caption{Convergence with the average}
\label{fig:average}
\end{figure}

Figure~\ref{fig:average} shows the results for learners in the version
of the game where $y$ is the average strategy of other agents.  Each
curve shows the distance from equilibrium as a function of the
number of rounds of a population of agents of
a given size using a given learning algorithm.  The results were
averaged over 10 runs.
Since the payoffs for nearby strategies are close,
we want our notion of distance to take into account that agents
playing 7 are closer to equilibrium (8) than those playing
zero.
Therefore, we consider the expected distance of $\rho$ from equilibrium:
$\sum_{a} \rho(a) |a - 8|$.  To determine $\rho$, we counted
the number of times each action was over
the length of a stage,
so in practice
the distance will never be zero due to mistakes and exploration.
For ease of presentation,
the graph shows only populations of size up to 100;
similar results were obtained for populations up to 5000 agents.

For stage learning, increasing the population size has a dramatic
impact.  With two agents, mistakes and best replies to the results of
these mistakes cause behavior to be quite chaotic.  With
ten agents, agents successfully learn, although mistakes and
suboptimal strategies are quite frequent.  With one hundred
agents, all the agents converge quickly to equilibrium strategies
and mistakes are rare; almost all of the distance from
equilibrium is due to exploration.

No-regret learning also converges quickly, but the ``quality'' of
convergence (how close we get to equilibrium) is not as high.
The major problem is that a significant fraction of agents play
near-optimal actions rather than optimal action.  This may have a
number of causes.
First, the guarantee is that the
asymptotic value of $\rho$ will be an equilibrium, which allows
the short periods that we consider to be far from equilibrium.
Second,
the quality of convergence depends on $\epsilon$,
so tight convergence may require a much lower rate of
exploration and thus a much longer convergence time.  Finally,
this algorithm is guaranteed to converge
only
to a correlated equilibrium,
which may not be a Nash equilibrium.

\begin{figure}
\centering
\includegraphics[height=2.25in]{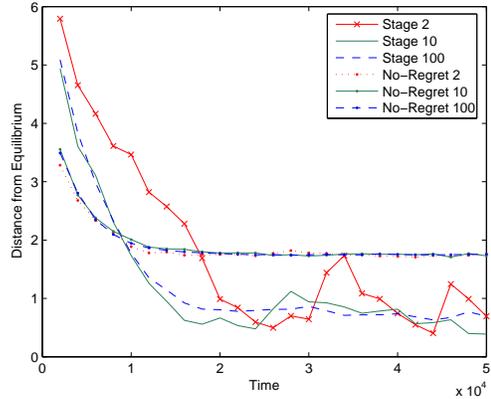}
\caption{Convergence with random matching}
\label{fig:matching}
\end{figure}

Figure~\ref{fig:matching} shows the results when agent payoffs are
determined by randomly matching agents.  Even for large numbers of
stage learners, convergence is not as tight and takes on the order of
ten times longer.  This is a result of the information available to
agents.  When payoffs were determined by the average strategy, a
single observation was sufficient to evaluate a strategy, so we could
use very short stages.  To deal with the noise introduced by random
matching we need much longer stages.  The number of stages to
convergence is similar.  Even with longer stages and a large number
of agents, mistakes are quite common.  Nevertheless agents do
successfully learn.  The performance of no-regret
learners is less affected because they use
payoff information from the entire run of the game,
while stage learners
discard payoff information at the end of each stage.

Convergence in the random-matching game takes approximately
20,000 rounds, which is too slow for many applications.  If a
system
design requires this type of matching, this makes learning problematic.
However, the results of Figure~\ref{fig:average} suggest that the
learning could be done much faster if the system designer could supply
agents with more information.  This suggests that collecting statistical
information
about the behavior of agents may be a critical feature for ensuring fast
convergence.
If agents know enough about the game to determine their
expected payoffs from this statistical information, then they can
directly learn, as in Figure~\ref{fig:average}.  Even with less
knowledge about the game, statistical information can still speed
learning, for
example, by helping an agent determine whether the results of exploring
an action were
typical or due to the other agent using a rare action.

\section{Discussion} \label{sec:discussion}
While our results show that a natural learning algorithm can learn efficiently
in an interesting class of games, there are many further issues that merit exploration.

\subsubsection*{Other Learning Algorithms}

Our theorem assumes that agents use a simple rule for learing within
each stage: they average the value of payoffs received.  However,
there are certainly other rules for estimating the
value of an action; any of these can be used as long as the rule
guarantees
that errors can be made arbitrarily rare given sufficient time.
It is also not necessary to restrict agents to stage learning.  
Stage learning
guarantees a stationary
environment for a period of time, but such strict behavior
may not be needed or
practical.  Other approaches, such as exponentially
discounting the weight of observations~\cite{greenwald01,marden08} or
Win or Learn Fast~\cite{wolf}
allow an algorithm to focus its learning on recent observations
and provide
a stable environment
in which other agents can learn.

\subsubsection*{Other Update Rules}

In addition to using different algorithms to estimate the values of
actions, a learner could also change the way he uses those values to
update his behavior.  
For example, rather than basing his new strategy
on only the last stage, he could base it on the entire history of
stages and use a rule in the spirit of fictitious play.  Since there
are games where fictitious play converges but best-reply dynamics do
not, this could extend our results to another interesting class of
games, as long as the errors in each period do not accumulate over
time.  Another possibility is to update probabilistically or use a
tolerance to determine whether to update
(see e.g.~\cite{young06,hart01}).
This could allow convergence in games where
best-reply dynamics oscillate or decrease the fraction of agents who
make mistakes once the system reaches equilibrium.

\subsubsection*{Model Assumptions}

Our model makes several unrealistic assumptions, most notably that
there are countably many agents who all
share the
same utility function.
Essentially the same results holds with a large, finite number of
agents, adding a few more ``error terms''.  In particular, since there
is always a small probability that every agent makes a mistake at the
same time, we can prove only that no more
than a $1-e$ fraction of the agents make errors in most rounds, and that
agents spending most of their time playing equilibrium strategies.

We have also implicitly assumed that the set of agents is fixed.  We
could easily allow for \emph{churn}: agents entering and leaving the system.
A reasonable policy for newly-arriving
agents is to pick a random $a_\epsilon$ to use in the next stage.
If all agents do this, it follows that convergence is unaffected: we
can treat the new agents as part of the $e$ fraction that made a
mistake in the last stage.  Furthermore, this tells us that newly
arriving agents ``catch up'' very quickly.  After a single stage, new
agents are guaranteed to have learned a best reply with probability at
least $1-e$.
Finally,
we have assumed that all agents have the same utility function.
Our results can easily be extended to include a finite number of
different types of agents, each with their own utility function, since
the SLLN can be applied to each
type of agent.  We believe that our results hold even
if the set of possible types is infinite.  This can happen,
for example, if an agent's utility depends on a valuation drawn from
some interval.  However, some care is needed to define best-reply
sequences in this case.

\subsubsection*{State}

One common feature of distributed systems not addressed in this work
is state.
For example, in a scrip system where agents pay each
other for service using an internal currency or {\em scrip}, whether
an agent should seek to provide service depends on the amount of money
he currently has~\cite{scrip06}.

In principle, we could extend our
framework to games with state: in each stage each agent chooses a
policy to usually follow and explores other actions with probability
$\epsilon$.  Each agent could then use some \emph{off-policy algorithm}
(one where the agent can learn without controlling the
sequence of observations; see~\cite{kaebling96} for examples) to
learn an optimal policy to use in the next stage.
One major problem with this approach is that standard algorithms learn
too slowly for our purposes.  For example, Q-learning~\cite{watkins92}
typically needs to observe each state-action pair hundreds of times in
practice.
The low
exploration probability means that the expected $|S||A| / \epsilon$
rounds needed to explore each even once
for each
pair is large.
Efficient learning requires more
specialized algorithms that can make better use of the structure of a
problem, but this also makes providing a general guarantee of
convergence more difficult.
Another problem is that, even if an agent explores each action for
each of his possible local states, the payoff he receives will
depend on the states of the other agents and thus the actions they
chose.
We need
some property of the game to guarantees this distribution of
states is in some sense ``well behaved.''

Despite these concerns, preliminary results suggest that simple
learning algorithms work well for games with state.
In experiments on a game using the model of a scrip system
from~\cite{scrip06}, we found that a stage-learning algorithm
that uses a specialized algorithm for determining the value of actions
in each stage converges to equilibrium quickly despite churn and
agents learning at different rates.

\subsubsection*{Mixed Equilibria}

Another restriction of our results is that our agents only learn
pure strategies.  One way to address this is to discretize the mixed
strategy space (see e.g.~\cite{young06}).  If one of the resulting
strategies is sufficiently close to an equilibrium strategy and
best-reply dynamics converge with the discretized strategies, then
we expect agents to converge to a near-equilibrium distribution of
strategies. We have had empirical success using this approach to
learn to play rock-paper-scissors.

\subsubsection*{Unexpected and Byzantine Behavior}

In practice, we expect that not all agents will be trying to
learn optimal behavior in a large system.  Some agents may
simply play some particular (possibly mixed) strategy
that
they are
comfortable with, without trying to learn a better strategy.
Others may be learning but with an unanticipated
utility function.  Whatever their reasons, if these sufficiently few
such agents are choosing their strategies i.i.d. from fixed
distribtions (or at least fixed for each stage), then our results hold
without change.  This is because we already allow an $e$ fraction of
agents to make arbitrary mistakes, so we can treat these agents as
simply mistaken.

Byzantine agents, who might wish to disrupt learning as much as
possible, do not fit as neatly into our framework; they
need not play the same strategy for an entire stage.
However, we expect that since correct agents
are randomizing their decisions, a small number of Byzantine agents
should not be able to cause many agents to make mistakes.

\section{Conclusion} \label{sec:conclusion}

Learning in distributed systems requires algorithms that are scalable
to thousands of agents and can be implemented with minimal information
about the actions of other agents.  Most general-purpose multiagent
learning algorithms fail one or both of these requirements.
We have shown here that stage learning can be an efficient
solution
in large anonymous
games where approximate best-reply dynamics lead to approximate pure
strategy Nash equilibria.
Many interesting classes of games have this
property, and it is frequently found in designed games. In contrast
to previous work, the time to convergence guaranteed by the theorem
does not increase with the number of agents.
If system designers can find an appropriate game satisfying these
properties on which to base their systems, they can be confident that
nodes can efficiently learn appropriate behavior.

Our results also highlight two factors that aid convergence.  First,
having more learners
often
improves performance.  With more learners, the noise introduced
into payoffs by exploration and mistakes becomes more consistent.
Second, having more information
typically %
improves performance. %
Publicly available statistics about the observed behavior of agents %
can allow an agent to learn effectively
while making fewer local observations.

\subsection*{Acknowledgements}

EF, IK,
and JH are supported in part by NSF grant ITR-0325453.  JH is
also supported in part by NSF grant
IIS-0812045
and by AFOSR grants
FA9550-08-1-0438 and 
FA9550-05-1-0055.  EF is also supported in part by NSF grant
CDI-0835706.  

\bibliographystyle{abbrv}
\bibliography{Z:/Research/Bibliography/kash}

\end{document}